\documentstyle[12pt,epsf]{article}
\newlength{\extraspace}
\newlength{\extraspaces}
\newcommand{\be}{\begin{equation}
\addtolength{\abovedisplayskip}{\extraspaces}
\addtolength{\belowdisplayskip}{\extraspaces}
\addtolength{\abovedisplayshortskip}{\extraspace}
\addtolength{\belowdisplayshortskip}{\extraspace}}
\newcommand{\ee}{\end{equation}}
 
\newcommand{\ba}{\begin{eqnarray}
\addtolength{\abovedisplayskip}{\extraspaces}
\addtolength{\belowdisplayskip}{\extraspaces}
\addtolength{\abovedisplayshortskip}{\extraspace}
\addtolength{\belowdisplayshortskip}{\extraspace}}
\newcommand{\ea}{\end{eqnarray}}

\newcommand{\bas}{\begin{eqnarray*}
\addtolength{\abovedisplayskip}{\extraspaces}
\addtolength{\belowdisplayskip}{\extraspaces}
\addtolength{\abovedisplayshortskip}{\extraspace}
\addtolength{\belowdisplayshortskip}{\extraspace}}
\newcommand{\eas}{\end{eqnarray*}}

\newcounter{subequation}[equation]
\makeatletter

\expandafter\let\expandafter
\reset@font\csname reset@font\endcsname

\def\subeqnarray{\arraycolsep1pt
    \def\@eqnnum\stepcounter##1{\stepcounter{subequation}%
        {\reset@font\rm(\theequation\alph{subequation})}}
\jot5mm     \eqnarray}

\makeatother

\newcommand{\NP}[1]{Nucl.\ Phys.\ {\bf #1}}
\newcommand{\PL}[1]{Phys.\ Lett.\ {\bf #1}}

\newcommand{\bra}{\langle}
\newcommand{\ket}{\rangle}
\newcommand{\ra}{\rightarrow}

\newcommand{\nonum}{\nonumber \\[1.5mm]}
\newcommand{\sspace}{\makebox[1cm]{ }}

\newcommand{\nspace}{\!\!\!\!\!\!\!\!\!\!}

\newcommand{\lsim}{ \stackrel{\textstyle{<}}{\sim} }
\newcommand{\gsim}{ \stackrel{\textstyle{>}}{\sim} }

\newcommand{\lb}{\lambda}

\newcommand{\cA}{{\cal A}}

\newcommand{\cO}{{\cal O}}

\begin{document}
%
\begin{titlepage}
%
\renewcommand{\thefootnote}{\fnsymbol{footnote}}
\begin{flushright}
MPI-PhT/96-131\\
\end{flushright}
\vspace{1cm}

\begin{center}
{\LARGE A Scaling Hypothesis for the Spectral Densities\\[3mm]
In the O$(3)$ Nonlinear Sigma-Model}
\vspace{2cm}
 
{\large J. Balog\footnote{On leave of absence from the Research 
Institute for Particle and Nuclear Physics, Budapest, Hungary} and
M. Niedermaier}\\ [3mm]
{\small\sl Max-Planck-Institut f\"{u}r Physik}\\
{(\small\sl Werner Heisenberg Institut)}\\
{\small\sl F\"{o}hringer Ring 6, 80805 Munich, Germany}
\vspace{2cm}

{\bf Abstract}
\end{center}

\begin{quote}
A scaling hypothesis for the $n$-particle spectral densities of
the O$(3)$ nonlinear sigma-model is described. It states that 
for large particle numbers the $n$-particle spectral densities
are ``self-similar'' in being basically rescaled copies of a 
universal shape function. This can be viewed as a 2-dimensional,
but non-perturbative analogue of the KNO scaling in QCD. 
Promoted to a working hypothesis, it allows one to compute the 
two point functions at {\em all} energy or length scales.
In addition, the values of two non-perturbative constants
(needed for a parameter-free matching of the perturbative and 
the non-perturbative regime) are determined exactly.  
\end{quote}
\vfill
\renewcommand{\thefootnote}{\arabic{footnote}}
\setcounter{footnote}{0}
\end{titlepage}

{\em 1. Introduction:}
An efficient way to describe the two-point function of some local
operator $\cO$ in a relativistic QFT is in terms of a K\"{a}llen-Lehmann 
spectral representation. The spectral density $\rho(\mu)$ of $\cO$ 
can be viewed as a measure for the number of degrees of freedom 
coupling to $\cO$ at energy $\mu$. It decomposes into a sum of $n$-particle 
contributions 
\be
\rho(\mu) = \sum_{n} \rho^{(n)}(\mu)\;,
\label{spec1}
\ee
where, depending on the local operator under consideration, some of
the $n$-particle contributions may vanish on the grounds of internal
quantum numbers. In a theory with a single mass scale $m$ one has 
$\rho^{(1)}(\mu) \sim \delta(\mu -m)$ and $\rho^{(n)}(\mu),\;n\geq 2$ 
has support only above $n m$, i.e. above the $n$-particle threshold. 
Once $\rho(\mu)$ is known, the various (Minkowski space or Euclidean)
two-point functions of $\cO$ can be computed as convolution integrals 
with an appropriate kernel carrying only kinematical information. 
The dynamical problem consists in computing the $n$-particle spectral 
densities $\rho^{(n)}(\mu)$ of $\cO$.

Here we shall be concerned with massive 1+1 dimensional QFTs and 
in particular with the O$(3)$ nonlinear sigma (NLS) model. 
The four most interesting local operators in this model are:
The spin field, the Noether current, the energy momentum (EM) 
tensor and the topological charge (TC) density. Their spectral 
densities can be grouped into two families $\rho_l^{(n)}(\mu),
\;n\geq 1,\;l=0,1$ according to their isospin. For $n$ even/odd the 
$\rho_0^{(n)}$ are the spectral densities of the EM-tensor/TC-density,
respectively; similarly $\rho_1^{(n)}$ for $n$ even/odd are the 
spectral densities of the
Current/Spin, respectively. The following pieces of information are
available for these spectral densities: (i) For small particle
numbers $n$ the functions $\rho_l^{(n)}(\mu)$ can be computed 
{\em exactly} by means of the form factor approach. In \cite{BN}
this has been done up to 6 particles. (ii) For all particle 
numbers $n$ the $\mu \ra \infty$ asymptotics of the $n$-particle 
spectral densities is known, and is given by 
\be 
\rho^{(n)}_l(\mu) \,\sim\, \frac{A^{(n)}_l}{\mu (\ln\mu)^{4-2l}}\;,
\sspace \mu \ra \infty\;,
\label{asy1}
\ee
where the constants $A^{(n)}_l$ are computable from the integrals 
of the lower particle spectral densities \cite{BN}.
The constants $A_l^{(n)}$ are rapidly increasing with $n$. This 
implies that the $\mu \ra \infty$ asymptotics of the full spectral
densities (1) cannot be computed by naively summing up the 
asymptotic expressions (\ref{asy1}), which in fact would be 
divergent. (iii) The large $\mu$ asymptotics of the full spectral
densities can however be computed in renormalized perturbation theory
(PT).%
\footnote{The correctness of PT in this model has been challenged in
\cite{PS}. To simplify the exposition we shall assume the validity
of PT for the UV asymptotics throughout this letter.}
One finds for the leading behavior 
\ba
\!\!\mbox{EM \& top:}&\nspace\;\,& 
\rho(\mu)\sim\frac{A^{\cO}}{\mu}\left[\frac{1}{(\ln\mu)^2} +
O\left(\frac{\ln\ln\mu}{(\ln\mu)^3}\right)\right],\nonum
\mbox{spin \& curr:}&\nspace\;\,& 
\rho(\mu)\sim\frac{A^{\cO}}{\mu}\left[1+ O\left(
\frac{1}{\ln \mu}\right)\right]\,.
\label{asy3}
\ea
Subleading terms can also be computed, but not all of the overall
constants are accessible to PT. In particular $\lb_1 := A^{\rm spin}$ 
is an unkown non-perturbative constant. In the case of the TC density 
$A^{\rm top}$ is fixed by PT but its relation to the non-perturbatively 
defined spectral sum (1) is not. Equivalently the matrix elements of
the TC density between the vacuum and some multi-particle 
state are defined only up to an unknown non-perturbative constant
$\lb_0$.

Missing pieces of information about the spectral densities are:
(iv) One would like to be able to compute the full spectral densities
for all $\mu \geq 0$, not only their large $\mu$ asymptotics.
This would allow one to compute the two-point functions at  
{\em all} energy/length scales. In terms of the spectral resolution 
(1) this amounts to knowing all the $n$-particle contributions, not 
only those with $n \leq n_0$ for which the computation can be done 
explicitly.   
(v) One would like to know the (exact) values of the non-perturbative
constants $\lb_0$ and $\lb_1$. Knowledge of these constants would allow 
one to match non-perturbative and perturbative information 
unambiguously. In this respect their role is similar to that of the 
$m/\Lambda$ ratio \cite{HN}.  

The purpose of this letter is to bridge the gap between the 
perturbative and the non-perturbative regime and to provide the 
missing pieces of information (iv) and (v). It is based on
a remarkable self-similarity property of the $n$-particle 
spectral densities. For large $n$ they appear to be basically 
rescaled copies of a ``universal shape function'' $Y_l(z)$. 
Explicitly
\be
\rho^{(n)}_l(\mu) \approx \frac{M_l^{(n)}}{\mu}\, Y_l\left(
\frac{\ln (\mu/m)}{\xi_l^{(n)}}\right)\;,\;\;l=0,1\,,
\label{scal1}
\ee
where $M_l^{(n)}$ and $\xi_l^{(n)}$ are certain scaling parameters
to be specified later. In the following we shall first give a 
precise formulation of the scaling law (\ref{scal1}) and recall some 
of the evidence presented for it in \cite{BN}. Then we  shall promote 
it to a working hypothesis and show that it has the following 
consequences: The UV behavior is consistent with PT; in particular 
those coefficients $A^{\cO}$ in (\ref{asy3}) accessible to PT are 
reproduced. The non-perturbative constants $\lb_0$ and $\lb_1$ are 
determined exactly and in the normalization \cite{BN} are given by 
\be
\lb_0 = \frac{1}{4}\;,\;\;\; \lb_1 = \frac{4}{3 \pi^2}\;.
\label{lb}
\ee
Finally candidate results for the two-point functions at {\em all}
energy or length scales are obtained.

\begin{figure}[htb]
\begin{flushleft}
\leavevmode
\epsfxsize=170mm
\epsfysize=100mm
\epsfbox{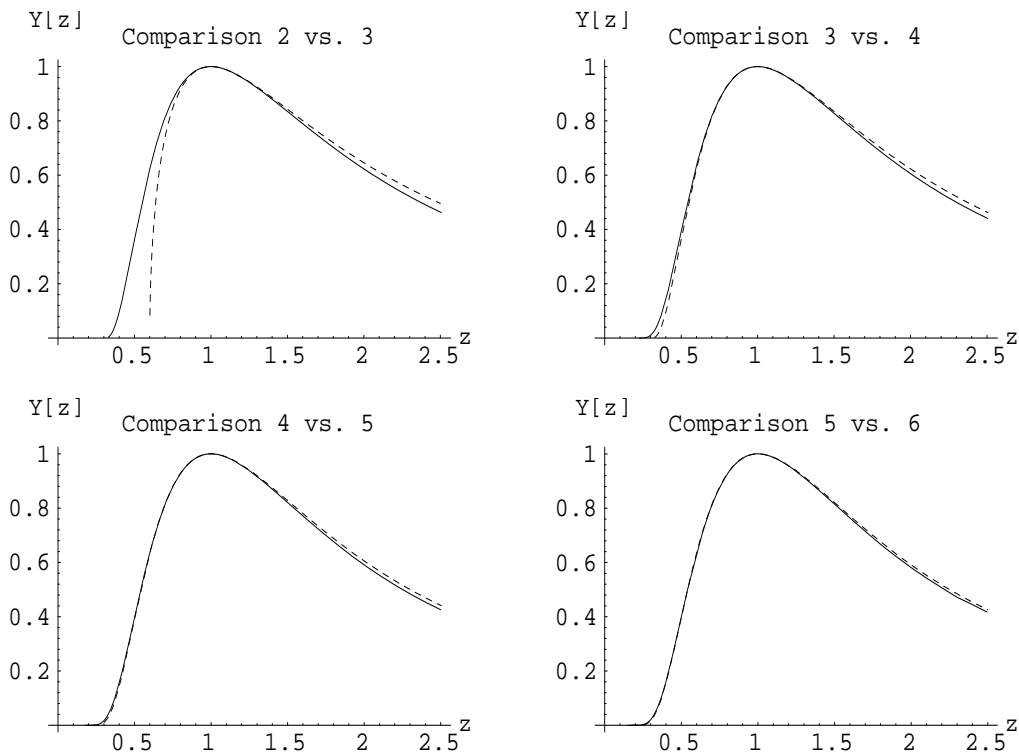}
\vskip -6mm
\end{flushleft}
\caption{Illustration of the self-similarity property of the rescaled $l=1$ 
spectral densities. The plots show $Y_1^{(n)}(z)$ (dashed) 
compared with $Y_1^{(n+1)}(z)$ (solid) for $n=2,3,4,5$.}
\label{SScurr}
\end{figure}

{\em 2. Formulation of the Hypothesis:}
With hindsight to the asymptotics (\ref{asy1}) let us introduce
\be
R^{(n)}_l(x) := m e^x \,\rho^{(n)}_l(m e^x)\,,\;\;l=0,1\,.
\label{scal2}
\ee
Here $l=0,1$ as before correspond to the EM tensor \& TC density and 
Spin \& Current series, respectively. The graphs of these functions 
are roughly `bell-shaped': Starting from zero at $x = \ln n$
they are strictly increasing,  reach a single maximum at some 
$x =\xi^{(n)}_l > \ln n$ and then decrease monotonically for all 
$x > \xi^{(n)}_l$. The position $\xi^{(n)}_l$ of the maximum
and its value $M^{(n)}_l = R^{(n)}_l(\xi^{(n)}_l)$ are two important
characteristics of the function, and hence of the spectral density.  
Defining
\be
Y^{(n)}_l(z) := \frac{1}{M^{(n)}_l}\,R^{(n)}_l(\xi^{(n)}_l z)\;,
\;\;l=0,1\,,
\label{scal3}
\ee
both the value and the position of the maximum are normalized to unity. 
Initially $Y^{(n)}_l(z)$ is defined for $(\ln n)/\xi^{(n)}_l\leq z<\infty$;
in order to have a common domain of definition we set $Y^{(n)}_l(z) 
= 0$ for $0\leq z\leq (\ln n)/\xi^{(n)}_l$. 
The proposed behavior of the spectral densities is as follows:
\vspace{2mm}

\noindent {\bf Scaling Hypothesis:}
\begin{itemize}
\item[(a)] {\em (Self-similarity) The functions $Y_l^{(n)}(z),\;n\geq 2$ 
converge pointwise to a bounded function $Y_l(z)$. The sequence of
$k$-th moments converges to the $k$-th moments of $Y_l(z)$ for $k+l=0,1$, 
i.e.}
\bas 
\nspace &&\lim_{n\ra \infty} Y^{(n)}_l(z) = Y_l(z)\;,\;\; z\geq 0\;,\nonum
\nspace && \lim_{n\ra \infty}\int_0^{\infty}dz\, z^k Y_l^{(n)}(z) =
\int_0^{\infty}dz\, z^k Y_l(z)\;.
\eas
\item[(b)] {\em (Asymptotic scaling) The parameters $\xi^{(n)}_l$ and 
$M^{(n)}_l$ scale asymptotically according to powers of $n$, i.e.}
$$
\xi_l^{(n)} \sim \xi_l\,n^{1+ \alpha_l}\;,\sspace
M_l^{(n)} \sim M_l\,n^{-\gamma_l}\;. 
$$
\end{itemize}
Feature (a) in particular means that for sufficiently large $n$ the 
graphs of two subsequent members $Y^{(n-1)}_l(z)$ and $Y^{(n)}_l(z)$
should become practically indistinguishable. This appears to be 
satisfied remarkably well even for small $n=4,5,6$, as is 
illustrated in Figure \ref{SScurr} for the $l=1$ series.

The analysis of part (b) of the scaling hypothesis is more involved.
It turns out that all but one of the exponents in (b) are fixed by 
self-consistency, and only this one has to be determined by fitting 
against the $n\leq 6$ particle data. The result is \cite{BN} 
\ba 
\gamma_1 =1\;, && \alpha_0 =\alpha_1 =:\alpha\;,\nonum
\gamma_0 = 3 +2\alpha\;,&& \alpha \approx 0.273\;.
\label{exponents}
\ea
{\em 3. Consequences of the Hypothesis:}
Let $n_0$ be the maximal particle number for which the spectral densities
have been computed explicitly (at present $n_0=6$). Then (\ref{scal1})
gives candidate expressions for all $n > n_0$ particle spectral densities
so that one can evaluate their sum. Concerning the UV behavior of the 
sum notice that a finite number of terms in the sum, each decaying
according to (\ref{asy1}), can never produce the different UV behavior
in (\ref{asy3}). However the infinite sum does. What is happening is 
that the partial sums $(\ln \mu)^{2 -2 l} \sum_{n_0+1}^N \rho_l^{(n)}(\mu)$
develop a plateau, i.e. are practically constant in a large interval
$\mu_{min} \lsim \mu \lsim \mu_{max}(N)$. When $N$ is increased the 
plateau is prolonged and eventually reaches out to 
infinity, i.e. $\mu_{max}(N) \ra \infty$ for $N \ra \infty$ (while 
$\mu_{min}$ is basically $N$-independent). The value of the plateau
determines the asymptotic constants in (\ref{asy3}). In those cases 
where the coefficients are accessible to PT the perturbative value
is reproduced with an accuracy better than 1\%. In addition one obtains 
the following two {\em exact} relations among the four constants 
\be
A^{\rm curr} = \frac{\pi}{4} A^{\rm spin}\;,\;\;\;
A^{\rm EM} = 4 A^{\rm top}\;. 
\label{c1}
\ee
These relations reflect the linking between the even and the odd 
particle members of an isospin family, which results from the 
clustering relations obeyed by the exact form factors\cite{BN}.   
Since $(A^{\rm curr})_{PT} = 1/3\pi$ the first eq. in (\ref{c1})
gives $A^{\rm spin} =\lb_1$ as in (\ref{lb}), while the second eq.    
fixes the physical normalization of the TC matrix elements in terms 
of that of the EM tensor, which in our conventions amounts to 
$\lb_0 =1/4$. 

\begin{figure}[htb]
\begin{center}
\vskip 2mm
\leavevmode
\epsfxsize=130mm
\epsfysize=60mm
\epsfbox{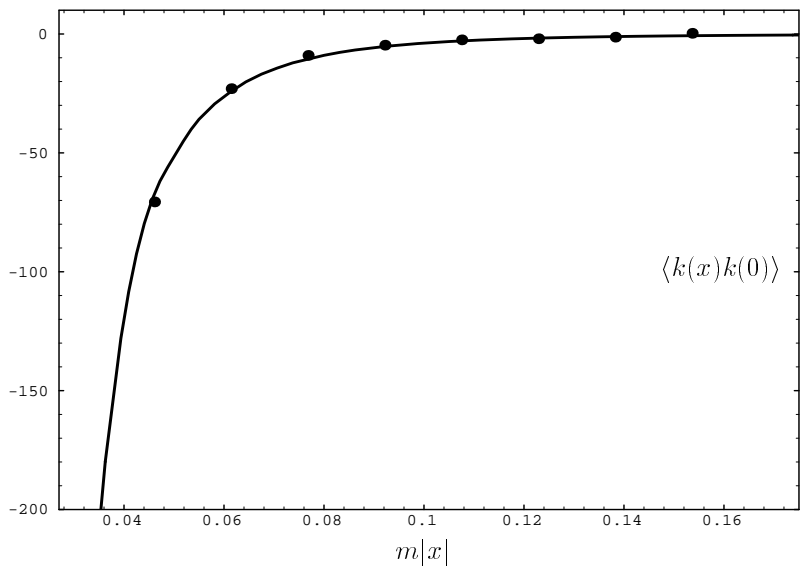}
\vskip 5mm
\end{center}
\caption{Two-point function of the topological charge density.
Comparison MC data and form factor result. Test of $\lb_0 =1/4$.}
\label{TCplot}
\end{figure}

The $\lb_0$ value can be tested independently by 
means of MC simulations. In Figure \ref{TCplot} the results for the 
two-point functions of the TC density is shown
-- once computed via the form factor approach, with the absolute
normalization fixed according to (\ref{lb}) and once via MC simulations.
The simulations were done using the cluster algorithm of \cite{TC} with 
the standard action and the geometrical definition of the TC density. 
The data in Figure 2 correspond to a 460 square lattice at 
inverse coupling $\beta =1.80$ (correlation length $65.05$).
The statistical errors are smaller than the size of the dots. 
The nice agreement confirms $\lb_0 =1/4$ and hence provides further
support for the scaling hypothesis.   

The plateau-phenomenon for the spectral densities has a counterpart for
the two-point functions. In momentum space the latter can be computed as
Stieltjes transforms of the corresponding spectral densities. Inserting the 
decomposition (\ref{spec1}) one obtains for the $n\leq N$ particle 
approximants
\be
I_N(y) = \sum_1^N \int_0^{\infty}
d\mu\, \frac{\rho^{(n)}(\mu)}{\mu^2 + m^2 e^{2 y}}\;,
\;\;y = \ln p/m\,.
\label{Isum}
\ee
For $n > n_0$ the $n$-particle spectral densities are evaluated by means  
of the scaling hypothesis. The $N \ra \infty$ limit of (\ref{Isum}) yields  
a candidate for the exact two-point function. 

\begin{figure}[htb]
\begin{flushleft}
\vskip 3mm
\leavevmode
\epsfxsize=165mm
\epsfysize=80mm
\epsfbox{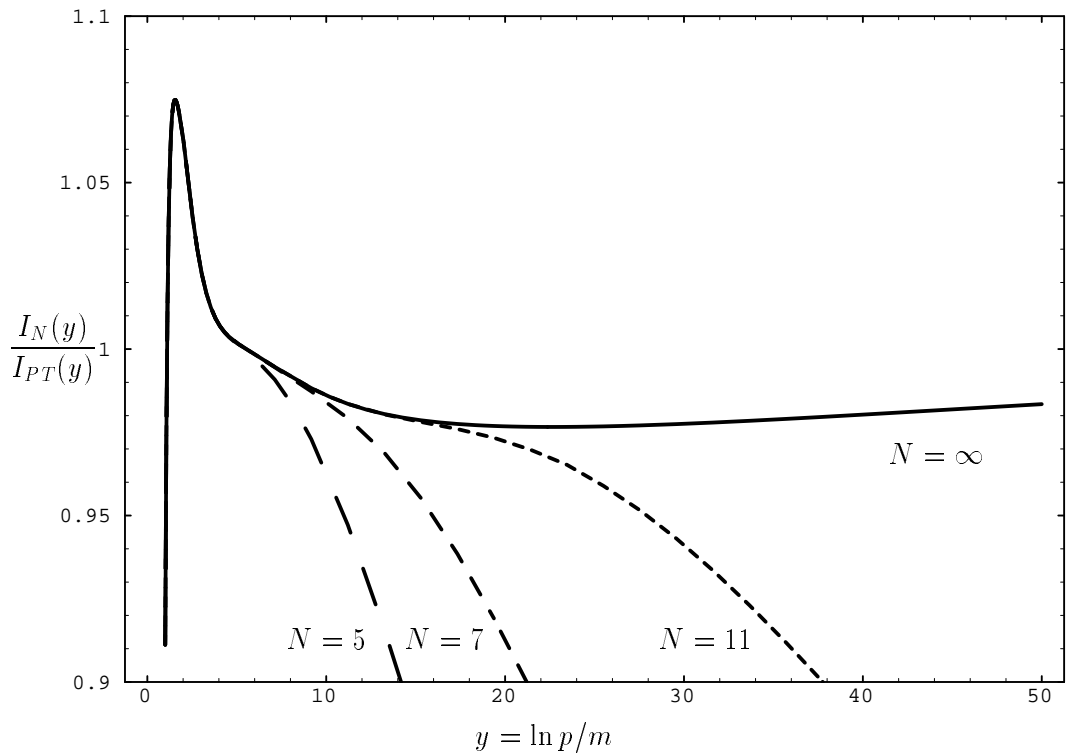}
\vskip 8mm
\end{flushleft}
\caption{Spin two-point function: Ratio of $I_N(y)$ approximants
and 2-loop perturbation theory $I_{PT}(y)$ versus $y= \ln p/m$.} 
\label{PJoddplot}
\end{figure}

For the case of the spin field the result is shown in Figure 
\ref{PJoddplot} in units of the 2-loop PT result 
for the same quantity. The `exact' $N =\infty$ curve approaches its 
asymptotic value (here normalized to unity by using 
$\lb_1 =4/(3 \pi^2)$ for the normalization of the PT result) fairly slowly. 
Demanding a $1\%$ accuracy for PT one finds that there are two such 
regimes, one at an intermediate energy scale $ 4 \lsim y \lsim 9$ 
and a second one for $y \gsim 75$. A similar behavior is found for 
the current two-point function. Thus, somewhat surprisingly, PT does 
not necessarily improve monotonically with increasing energy. 
As a reservation one must add however that these are $1\% -2\%$ effects 
which may be affected by subleading terms in the scaling law (b). 
(Inclusion of a 3-loop PT contribution does not remove the effect;
subleading terms arising from extrapolating $Y^{(n+1)}(z) - Y^{(n)}(z)$
have been taken into account in Figure 3.)

{\em 4. Relation to KNO Scaling:}
A bonus of the above scaling hypothesis is that it implies KNO-like 
scaling laws for multi-particle production processes (but not vice versa).
Of course there is no particle production in an integrable model, but
we can enlarge our model-world by a ``weak" sector so that a general state
will be  of the form $|s;w\ket$, where $s$ is the ``strong" (sigma-model) part
of the state and $w$ is the ``weak" part. Adding a current-current interaction
term, production processes $|s;w\ket\to|s^\prime;w^\prime\ket$
(where the states $s$ and $s^\prime$  have different particle
numbers) become possible. The corresponding transition amplitude, to
lowest order in the new interaction, reads
\be
\cA=\int\, d^2x\,l^{a\mu}(x)\,\bra s^\prime|j^a_\mu(x)|s\ket\,,
\label{amplitude}
\ee
where the Fourier transform of $l^a_\mu(x)$ has support at the
momentum $Q$, the ``weak" momentum transfer.

The simplest production process is the two-dimensional analogue of the
$e^+e^-$ annihilation. We model this process by choosing $Q^2>0$ and
$s=|0\ket$. Summing over all (discrete and continuous) quantum numbers 
of the $n$-particle state $s^\prime$, the probability
distribution for producing $n$ particles at energy $\mu=\sqrt{Q^2}$
becomes independent of the ``weak" part of the process and is proportional
to the current spectral density. The proportionality factor involves 
the coupling constant of the external current-current interaction,
which drops out when considering the conditional probability $P_{2 n}$ 
for having exactly $2 n$ particles produced, once the process has taken 
place at all. One has 
\be
P_{2n}= \frac{\mu}{\kappa}\rho_1^{(2n)}(\mu)\,,\sspace 
\sum_{n=1}^{\infty}P_{2 n} =1\,,
\label{distribution}
\ee
where the second eq. fixes the normalization constant $\kappa$.
Similarly production processes upon perturbation with the EM-tensor 
can be studied, in which case $\rho_0^{(2n)}(\mu)$ appears in 
(\ref{distribution}). Using our scaling hypothesis the energy dependence
of the probability distribution can in both cases be expressed, for large 
$\mu$ (where one can approximate the sums by integrals), in terms of 
$\bar n= \sum_{n=1}^\infty\, 2n\,P_{2n}\sim (\ln \mu)^{1/(1+\alpha)}$, the 
average number of particles produced. The asymptotic distribution takes 
the KNO-scaling form \cite{KNO}
\be
\bar n\,P_{2n}=f\left(\frac{n}{\bar n}\right)\,.
\label{KNOform}
\ee
The scaling function $f(q)$ is given in terms of the
universal shape function as
\be
f(q)=\frac{2(1+\alpha)\tilde{h}_l}{h_l^2}
\left(\frac{h_l}{2\tilde{h}_lq}\right)^{\gamma_l}\;Y_l
\left(\left(\frac{h_l}{2\tilde h_l\,q}\right)^{1+\alpha}\right)\,,
\label{scalingfc}
\ee
where the parameters $h_l$ and $\tilde h_l$ are the 
$\big(\frac{\gamma_l-1}{1+\alpha} -1\big)^{\rm th}$ and
$\big(\frac{\gamma_l-2}{1+\alpha} -1\big)^{\rm th}$ moments of the 
universal shape function $Y_l(z)$, respectively, and the exponents are
given in (\ref{exponents}). The case $l=1$ 
corresponds to the current perturbation and $l=0$ to the perturbation
by the EM-tensor. These KNO-type scaling laws, however, are only valid 
for simultaneously large particle numbers and large energies. In contrast,
the scaling hypothesis (a), (b) for the spectral densities is valid for all
energies, and in particular is non-perturbative in nature.

\vspace{5mm}
{\tt Acknowledgements:} We are indebted to F. Niedermayer for 
making available to us his MC program. We also wish to thank
P. Weisz for discussions and W. Ochs and S. Lupia for calling our 
attention to the KNO scaling. M.N. acknowledges support by the 
Reimar L\"{u}st fellowship of the Max-Planck-Society.  

%


\vspace{5mm}

\begin{thebibliography}{10}
\bibitem{BN} J. Balog and M. Niedermaier, MPI-Ph/96-120; hep-th/9612039.


\bibitem{HN} P. Hasenfratz, M. Maggiore
and F. Niedermayer, \PL{B245} (1990) 522. 

\bibitem{TC} M. Blatter, R. Burkhalter, P. Hasenfratz and F. Niedermayer,
Phys. Rev. {\bf D53} (1996) 923.

\bibitem{PS} 
A. Patrascioiu and E. Seiler, Phys. Rev. Lett {\bf 74} (1995) 1924.

\bibitem{KNO} Z. Koba, H.B. Nielsen and P. Olesen, \NP{B40} (1972) 317.\\
              A.M. Polyakov, Sov. Phys. JETP {\bf 32} (1971) 296; {\bf 33}
              (1971) 850.
\end{thebibliography}
\end{document}